%% file: cartolabe.tex
\documentclass{IEEEcsmag}

\usepackage[utf8]{inputenc}

\ifCLASSOPTIONcompsoc
  \usepackage[nocompress]{cite}
\else
  \usepackage{cite}
\fi
\ifCLASSINFOpdf
  \graphicspath{{./figures/}}
  \DeclareGraphicsExtensions{.pdf,.jpeg,.png}
\else
  \graphicspath{{./figures/}}
  \DeclareGraphicsExtensions{.pdf, png,.eps}
\fi

\usepackage{amsmath}
\usepackage{amssymb}
\usepackage{mathptmx}
\usepackage{algorithmic}
\usepackage{multirow}

\ifCLASSOPTIONcompsoc
  \usepackage[caption=false,font=footnotesize,labelfont=sf,textfont=sf]{subfig}
\else
  \usepackage[caption=false,font=footnotesize]{subfig}
\fi
\usepackage[colorlinks,urlcolor=darkgray,linkcolor=darkgray,citecolor=darkgray]{hyperref}
\usepackage{upmath}
\usepackage{xspace}
\usepackage{xcolor}
\usepackage{comment}

\newif\ifcga \cgatrue 

\newcommand{\jdf}[1]{\textcolor{blue}{{\tiny jdf:}#1}}

\newcommand{\kmeans}{\textit{k}-means\xspace}

\newcommand{\RR}{\ensuremath{\mathbb{R}}}
\newcommand{\http}[1]{\href{https://#1}{#1}}

\def\CarD{\textsc{\sc Cartolabe-Data}}
\def\CarV{\textsc{Cartolabe-Vis}}
\def\Car{\textsc{\sc Cartolabe}}
\def\UMAP{\textsc{\sc Umap}}
\newcommand{\ts}{\textsuperscript}

\newcommand{\revv}[1]{#1}
\newcommand{\dell}[1]{}

\newcommand{\rev}[1]{#1}
\newenvironment{revs}{\bgroup}{\egroup} 
\newcommand{\del}[1]{}
\newcommand{\todo}[1]{}

\jvol{XX}
\jnum{XX}
\paper{8}
\jmonth{March}
\jname{Cartolabe}
\pubyear{2020}

\begin{document}

\sptitle{Department: Head}
\editor{Editor: Name, xxxx@email}

\title{Cartolabe: A Web-Based Scalable Visualization of \texorpdfstring{\\}{ }Large Document Collections}

\author{Philippe~Caillou}
\affil{Université Paris-Saclay, CNRS, Inria, LRI, France}

\author{Jonas~Renault}
\affil{Université Paris-Saclay, CNRS, LRI, France}

\author{Jean-Daniel~Fekete,~\IEEEmembership{Senior~Member,~IEEE}}
\affil{Université Paris-Saclay, CNRS, Inria, LRI, France}

\author{Anne-Catherine~Letournel}
\affil{Université Paris-Saclay, CNRS, LRI, France}

\author{Michèle~Sebag}
\affil{Université Paris-Saclay, CNRS, Inria, LRI, France}


\markboth{IEEE Computer Graphics and Applications}%
{Caillou \MakeLowercase{\textit{et al.}}: Cartolabe}


\input{00-abstract}

\maketitle

\IEEEpeerreviewmaketitle

\begin{figure*}
  \centering
  \includegraphics[width=\linewidth]{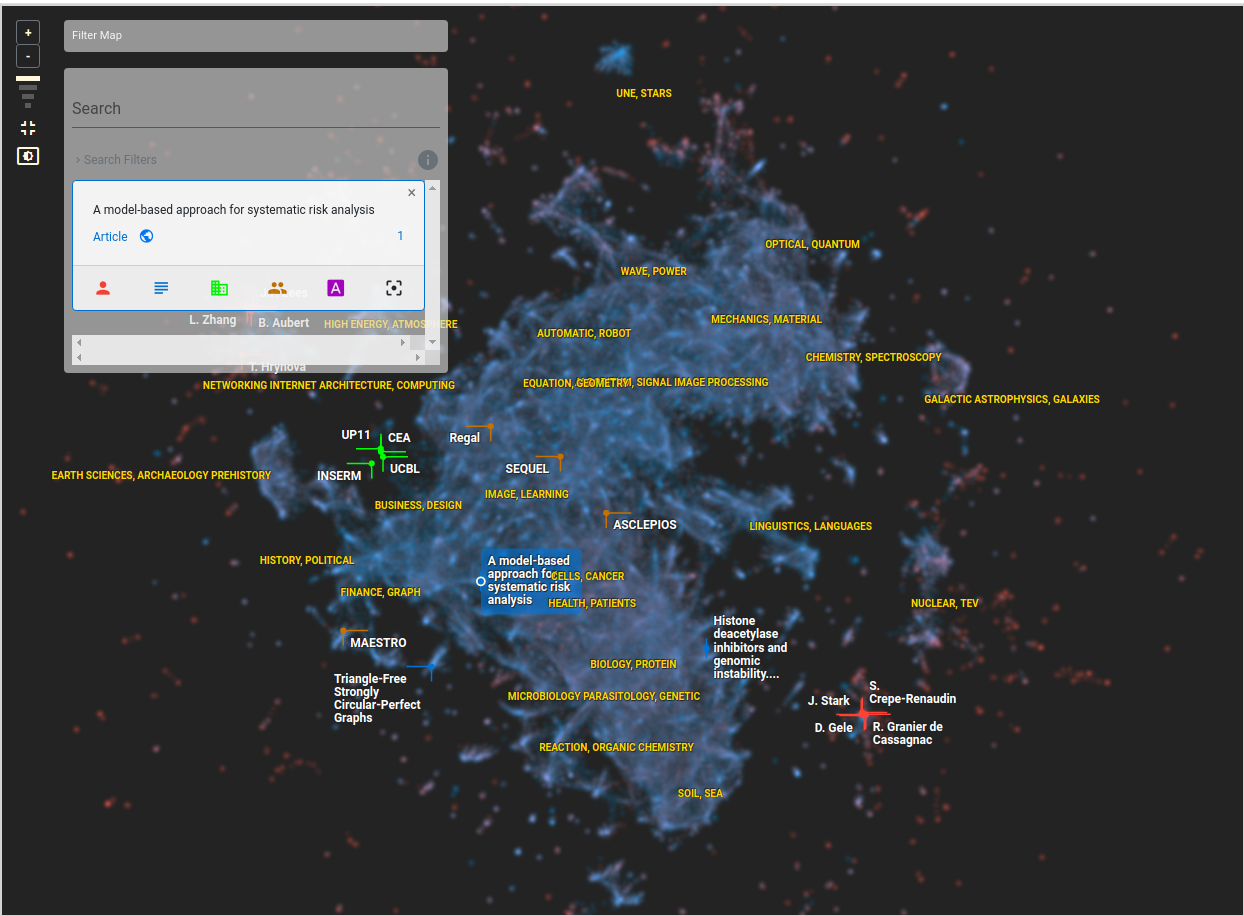}
  \caption{The \Car\ system: A Web-based, zoomable, 2D
    visualization of the HAL scientific publication repository, involving
    700k articles and 170k authors. This visualization is aimed at a large
    scientific audience with no particular knowledge in visualization. Topic headlines are displayed in yellow; important articles and authors are displayed in white.}
  \label{fig:hal}
\end{figure*}

\input{01-intro.tex}

\input{02-background.tex}
\input{03-overview.tex}

\input{03-A-CartoData.tex}
\input{03-B-CartoVis.tex}
\input{06-use-cases.tex}

\input{07-discussion}

\input{08-conclusion.tex}



\bibliographystyle{IEEEtran}
\bibliography{IEEEabrv,cartolabe}

\begin{IEEEbiography}{Philippe Caillou}{\,}is associate professor at the CS department of Université Paris-Saclay and CNRS.
Contact him at Philippe.Caillou@lri.fr.
\end{IEEEbiography}

\begin{IEEEbiography}{Jonas Renault,}{\,}is Software Engineer at CNRS, part of the CS department of Université Paris-Saclay and CNRS. Contact
him at jonasrenault@gmail.com.
\end{IEEEbiography}

\begin{IEEEbiography}{Jean-Daniel Fekete,}{\,}is a Senior Research Scientist at Inria, France, head of the Aviz Project-Team dedicated to visualization and visual analytics, and part of the CS department of Université Paris-Saclay and CNRS. Contact
him at Jean-Daniel.Fekete@inria.fr.
\end{IEEEbiography}

\begin{IEEEbiography}{Anne-Catherine Letournel,}{\,}is a Research Engineer at the CS department of Université Paris-Saclay and CNRS, in charge of the software development service for the laboratory.
Contact her at Anne-Catherine.Letournel@lri.fr.
\end{IEEEbiography}

\begin{IEEEbiography}{Michèle Sebag,}{\,}is Senior Research Scientist at CNRS, France, head of the A\&O team at the CS department of Université Paris-Saclay and CNRS. Contact
her at Michele.Sebag@lri.fr.
\end{IEEEbiography}

\end{document}

%% file: 00-abstract.tex
\begin{abstract}
We describe \Car, a web-based multi-scale system for visualizing and exploring large textual corpora based on topics, introducing a novel mechanism for the progressive visualization of filtering queries.
Initially designed to represent and navigate through scientific publications in different disciplines, \Car\ has evolved to become a generic framework and accommodate various corpora, ranging from Wikipedia (4.5M entries) to the French National Debate (4.3M entries).
\Car\ is made of two modules: the first relies on Natural Language Processing methods, converting a corpus and its entities (documents, authors, concepts) into high-dimensional vectors, computing their projection on the 2D plane, and extracting meaningful labels for regions of the plane. 
The second module is a web-based visualization, displaying tiles computed from the multidimensional projection of the corpus using the \UMAP\ projection method. 
This visualization module aims at enabling users with no expertise in visualization and data analysis to get an overview of their corpus, and to interact with it: exploring, querying, filtering, panning and zooming on regions of semantic interest. 
Three use cases are discussed to illustrate \Car's versatility and ability to bring large scale textual corpus visualization and exploration to a wide audience. 
\end{abstract}

%% file: 01-intro.tex
\chapterinitial{Introduction}\label{sec:introduction}
The visualization of large collections of documents, or corpora, faces both computational challenges and sense-making issues. 
These challenges and issues are at the crossroad of Visualization (VIS) and Machine Learning (ML).
On the VIS side, the objective is to provide a suitable overview and enable the user to get a quick understanding of the data through an effective 2D representation, as well as to provide means to interact with this representation to explore it in a fluid and controlled manner. 
On the ML side, the objective is to achieve compressed representations of large corpora --- typically using dimensionality reduction techniques --- in a way that preserves meaningful arrangements of the material and is computationally efficient.  

This article presents a \del{new generic system}\rev{software framework} called \Car\ (\http{cartolabe.fr}) comprehensively addressing the above algorithmic challenges related to visualization, machine learning, and sense-making. This system is intended for several application domains, e.g., exploring the state of art in a scientific field, getting an overview of the expertise of an organization or a country through documents, and making sense of a large set of documents produced by citizens over a national debate. 
\revv{\Car\ addresses several types of needs for mainly three categories of users: i) Organizations interested in sharing large collections of documents, making them readily accessible and searchable online; ii) Experts interested in making sense of their corpus, and exploiting it for diverse information retrieval tasks; iii) Data scientists and NLP experts, interested in the fast assessment of their data summarization and querying pipelines.}

From a data compression perspective, \Car\ extensively relies on ML methods specifically dedicated to 
natural language processing (NLP)~\cite{lsa,lda} and on dimensionality reduction~\cite{Nonato2018} methods. 
From an interactive visualization perspective, \Car\ relies on the state of the art in web-based interactive visualizations, supporting both the general display of a data landscape and the dynamic exploration of the contextual information through querying, \rev{progressive} filtering, panning and zooming.

The combination of both VIS and ML expertise 
enables users to visualize collections of millions of documents and to interactively explore the data through progressive computation and visualization~\cite{Progressive}.
Furthermore, these functionalities are offered on the web, freeing the end-users from deploying or maintaining any infrastructure, storage, or computational resources. 

\Car\  was initially designed 
to handle a corpus of scientific publications from the HAL repository (\http{hal.archives-ouvertes.fr}), containing about 700k documents and 2M authors.  Its specificity is to rely on the textual information, as opposed to the other main systems relying on the authorship and citation graphs.  
This specificity was instrumental in extending \Car\ to other types of data collections, 
ranging from Wikipedia (4.5M documents) 
to {\em Le Grand D\'ebat} (4M documents), recording the citizens' contributions to the national debate organized by the French government in May 2019. 

\rev{Interactive systems offering map visualizations of large corpora, e.g., {\sc PaperScape} (\http{paperscape.org}), 
most generally use graph layout techniques to compute the document positions, emphasizing the social ties between articles through co-authors or citations. In contrast, \Car\ focuses on the document contents: topics or linguistic properties.
A topic map provides an intrinsic view of the scientific domains that is complementary to social ties-based maps and, for quite a few tasks, might be less biased. 
\dell{For example, for finding representative authors around a topic (e.g., for selecting reviewers), finding representative topics around authors (e.g., to understand the main thematic topics covered by an author), finding representative articles around a topic (e.g., for literature reviews).}
\revv{It can be used, for example, to find representative authors around a topic (e.g., to select reviewers), to find representative topics around authors (e.g., to understand the main thematic topics covered by an author), or to find representative articles around a topic (e.g., for literature reviews).}
Its filtering functionalities also allow \Car\ to support comparisons across organizations or years. 
We do not pretend \Car\ addresses all the issues related to document exploration and distant reading. The IN-SPIRE system~\cite{ThemeScapes} has clearly made the point that it was important to provide multiple visualizations and interfaces on document collections to fully grasp their multiple facets; \Car\ is one component in that landscape, with scalability to tens of millions of documents and higher flexibility in NLP methods for its preparation pipeline.}

\rev{\Car\ has become a generic system or a \emph{software framework}}, able to process and visualize a large variety of corpora.
It is made of modular building blocks, which can be replaced or customized depending on the different types of entities to visualize and explore. 

A particular usage of \Car\ is to support the visual assessment of new NLP methods, providing a fast and versatile interactive environment to visually explore the extracted concepts and their organization. The system, taking in charge the visualization and interaction parts, allows NLP experts to focus on the data processing methods and to assess the impact of new methods or hyper-parameter settings.

The main three contributions of \Car\ are: \emph{i)} to support web-based real-time visualization and filtering of large document collections, up to tens of millions of documents; \emph{ii)} to provide contextual information during the interactive navigation, allowing for scalable targeted search and filtering; and \emph{iii)} to provide a platform to test new NLP methods for computing document similarities and search. These contributions rely on the combination of VIS and ML techniques, constructing an annotated landscape and generating tiled density maps to support navigation, search, filtering, and exploration \rev{at scale}. 

\dell{
\Car\ addresses several types of needs: 
\begin{itemize}
    \item Organizations interested in sharing large collections of documents, making them readily accessible and searchable online; 
    \item Experts interested in making sense of their corpus, and exploiting it for diverse information retrieval tasks;
    \item Data scientists and NLP experts, interested in the fast assessment of their data summarization and querying pipelines.
\end{itemize}
}
\revv{The remainder of the paper will describe and exemplify \Car\ on the HAL use-case. After briefly discussing the state of the art, we give an overview of the \Car\ architecture and functionalities. Then, we describe the \CarD\ and the \CarV\ pipelines. Applications (Wikipedia and \emph{Le Grand Débat}) are discussed further in section ''Use Cases'' and a preliminary evaluation in the section ''Evaluation''.
}
\dell{The remainder of the paper will describe and exemplify \Car\ on the HAL use-case. Other applications (Wikipedia and \emph{Le Grand Débat}) are discussed further in section ''Use Cases''.}


%% file: 02-background.tex
\section{Related works}\label{sec:background}

\Car's inspiration can be traced back to the {\sc Galaxies} visualization~\cite{ThemeScapes}, still used in the IN-SPIRE system from Pacific Northwest National Laboratory.
{\sc Galaxies} builds a scatterplot view based on NLP and projection methods, displaying a few thousand documents according to their similarity.  Although the concept remains similar, \Car\ relies on modern NLP and projection methods that have progressed greatly in the last decades. As noted in the original article, multiple visualizations are needed to reveal different facets of a corpus, and \Car\ 
\dell{does not pretend to replace them all, but it does}
replace, extends, and enhances the Galaxies module in multiple ways: it is more scalable, uses a modern projection method, and can be tailored to use any NLP method available from Python.\dell{to extract meaningful topics from documents and compute a distance between them.}

According to the survey and taxonomy of text visualization tools by Kucher and Kerren~\cite{Kucher2015} \Car\ supports the analytic tasks ``Text Summarization / Topic Analysis / Entity Extraction'', the visualization tasks ``Overview'', ``Navigation / Exploration\revv{''}\dell{""}, and ``Cluster / Classification / Categorization'', in the domains of ``Scientific Articles / Papers'', ``Reviews / (Medical) Reports'', 
``Editorial Media''\revv{,}\ and others not listed. The data it represents are corpora represented as ``Networks'', implicitly through $k$-nearest neighbors computation. \Car\ is a ``2D'' visualization, representing items as ``Clouds / Galaxies'' with a ``Metric-dependent'' alignment. Compared to similar tools listed in the ``Text Visualization Browser'' (\http{textvis.lnu.se}), it can visualize large datasets and is not meant for one specific corpus but is meant to be tailored to different kinds of textual corpora. \Car\ is both an application and a framework to create corpus visualizations using the Galaxy metaphor\dell{and a series of applications to explore the HAL publication repository, arXiv, Wikipedia, and many other specialized corpora}.

Several recent visualizations for document corpora use a network layout approach: the document similarities rely on co-authorship or co-citation relationships and a graph layout algorithm computes the document positions based on their connections. 
{\sc Paperscape} (\http{paperscape.org}) offers a map representing about 1.6M articles from the arXiv repository, enabling its interactive exploration using a scrollable and zoomable tiled structure. The Maps of Computer Science ({\sc MoCS})~\cite{DBLP:conf/apvis/FriedK14} propose a representation of terms from research papers in the DBLP repository, extracting and visually displaying the topics similarity. Various heatmaps can be overlaid to visualize the profile of specific researchers or institutions; however the {\sc MoCS} system is static and limited in interactive functionalities such as filtering and searching.
Another project, {\sc VOSViewer}~\cite{vosviewer} aims to visualize bibliometric networks, representing both articles, authors and labs, and using NLP techniques to construct term similarity maps based on a relationship network.  Its scalability, as reported in the original article, seems to be limited to some thousands of items (as opposed to tens of millions for \Car).

Another trend is that of multidimensional projections (MDP), mapping \emph{high-dimensional} points (up to thousands or even
millions of dimensions) into \emph{low-dimensional} ones (typically 2D or 3D; only 2D is considered in this article). 
The last decade has seen many improvements in the
quality, applicability, and scalability of MDP techniques, as detailed in \del{a most} recent MDP surveys~\cite{Nonato2018, Espadoto2019}. 
A recent MDP approach, \UMAP~\cite{2018arXivUMAP}\rev{,} delivers projections with similar quality as those of rigorous mathematical grounding,
revealing meaningful global structures, offering the possibility of quickly projecting new points on a learned manifold and having good clustering properties. A main strength of \UMAP\ is its reasonable computational complexity.  

Contrasting with {\sc Paperscape} and {\sc VOSViewer}, \Car\ relies on the document contents to build their similarity, as opposed to co-authoring and co-citation graphs. This choice was instrumental to extend \Car\ to various document collections besides scientific articles. Contrasting with {\sc Galaxies} and {\sc MoCS}, \Car\ facilitates the user's exploration of data through an interactive map, with panning, zooming, and searching functionalities. Finally, \Car\ uses the \UMAP\ projection, projecting millions of points based on their similarity into a 2D map in a few hours on a standard desktop.

\rev{\Car\ shares many of the goals of the UTOPIAN system~\cite{UTOPIAN}, that is also meant to explore text corpora such as publications, using MDP techniques for visualization and NLP technique for topic analysis. Contrary to \Car, UTOPIAN can refine and change the NLP parameters interactively to tune the analysis and visualization as part of its exploration process. Providing interactive control over the NLP analysis is a very useful tool for application designers but not for \Car\ target users. \Car\ performs the NLP analyses offline and does not expose its pipeline to the final users\dell{, but it provides a more flexible pipeline than UTOPIAN for \Car\ application designers}\revv{. Whereas UTOPIAN relies on Nonnegative Matrix Factorization for topic modeling, \Car\ allows choosing among a large variety of modelings, including LSA, LDA, doc2vec, and any other implemented in Python, depending on the characteristics of the corpus}. Additionally, UTOPIAN is not designed to scale to millions of documents and is not web-based. 

\Car\ also shares many of the goals of the ``BioVis Explorer''~\cite{BioVisExplorer} that presents all publications in biological visualization using a MDP technique and allows filtering based on different features of its corpus. Compared to \Car, each article is presented as a thumbnail image in a 2D map, allowing an easy selection. However, while the BioVis Explorer visualizes about 150 articles, \Car\ is meant to scale to tens of millions and therefore should uses a heatmap instead of showing the individual points as thumbnail images. In addition, \Car\ shows both articles and authors over the same map by considering them both as bag of topics. It actually also considers higher-level entities such as teams, labs, and organizations as bag of topics too; \Car\ displays them on the same map accordingly.
}

\Car\ also shares the goal of {\sc Nanocubes}~\cite{nanocubes} to provide interactive visualization, searching, filtering, and aggregation capabilities over the web. Contrary to {\sc Nanocubes}, \Car\ uses progressive mechanisms~\cite{Progressive} to support these searching and filtering functionalities, simplifying its architecture while \del{exhibiting}\rev{sustaining} interactive latencies with large data over the web.

%% file: 03-overview.tex
\section{\Car}\label{sec:cartolabe}

\Car\ (\autoref{fig:hal}) is a web-based visualization and exploration system originally targeted at an audience of scientists of any discipline, to visualize their publications  arranged on the 2D plane according to topic contents. The only requirement on the target users is a general understanding of the considered corpus; no expertise in visualization or data analysis is assumed. 

\paragraph{Architecture overview}
\label{subsec:archi}
\Car\ comprises two modules that can be used independently: a data processing module and a visualization module respectively referred to as \CarD\ and 
\CarV. Both modules are publicly available (\http{gitlab.inria.fr/caillou/cartolabe-data} and \http{gitlab.inria.fr/caillou/cartolabe-visu})  and can be customized for various document collections. 

\CarD\ uses unsupervised text analysis to build a 2D representation of the document collection. This collection\rev{, typically retrieved by harvesting a site through a web API and transformed into a csv file for ingestion in \CarD}, is first processed to yield a compressed representation \del{(a JSON file)} including all information required for the visualization (entity labels, 2D coordinates, and nearest neighbors).
\CarV\ imports this \del{JSON} file and acts as a visualization server to display the map from web browsers, supporting its flexible exploration by the user. 

We first describe the functionalities of the system before detailing its architecture and modules.

\begin{figure}
  \centering
    \includegraphics[width=0.9\linewidth]{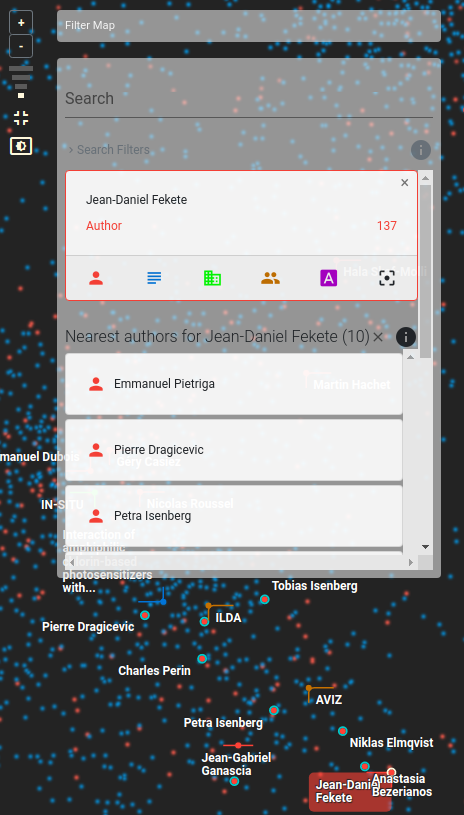}
\caption{Zooming on a specific point of interest: its semantic nearest neighbors  are displayed as dots with a cyan border.}
  \label{fig:sebag}
\end{figure}

\paragraph{Functionalities}
\label{subsec:overview}
\Car\ provides a general overview of the document collection (\autoref{fig:hal}). In the HAL application, the main entities are authors \rev{(visualized in red) and articles (blue)}. Both author and article maps are represented as heatmaps, with denser areas being brighter. These maps \rev{are blended by default but can be shown separately if the user is only interested in one of the entities}, and labels are displayed on top of these density maps to give contextual information and help users identify the specifics of a region:
\begin{itemize}
    \item Yellow labels characterize thematic regions. For instance, the cluster on the very right of the map includes articles and authors related to ``nuclear \& TEV''.
    \item White labels indicate the most \del{salient}\rev{important} entities in the map, depending on the zoom level. Their color background characterize the type of the entity (articles in blue, authors in red, laboratories in green and Inria ``Project Teams'' in brown).
\end{itemize}

This map conveys some organization of the data, displaying well-separated clusters and summarizing their themes (reminding that the textual analysis is unsupervised). 
Users can pan and zoom the map to get an \del{intuitive and precise idea }\rev{overview} of the data organization; the labels are refreshed accordingly to display the \del{salient entities at scale}\rev{important entities according to the scale and viewport}. 
Users can also select an item on the map and read the associated details on the search box on the top left of the screen, or search for an item based on its name and see where it is located on the map.

\paragraph{Search and Neighborhoods}
\label{subsec:neighbors}
A main functionality of \Car\ is to retrieve the entities (articles or authors) close to a given entity. This information retrieval ability supports different usages, such as finding experts to review a paper or a technical proposal, or finding papers relevant to a topic to complete a bibliographic study.
\def\XX{Jean-Daniel Fekete}
For instance, let us query an author (here \XX, visible in the bottom of the screen; \autoref{fig:sebag}). The most semantically related articles or authors are highlighted in the map and displayed in the search box. 

Note that these semantic nearest neighbors might be far from the location of the query, due to artifacts of the dimensionality reduction~\cite{Nonato2018}: entities naturally live in a much higher dimensional space than the 2D space. We shall come back to this issue later (Section ''Landmarks''). 

\begin{figure*}
  \centering
  \includegraphics[width=\linewidth]{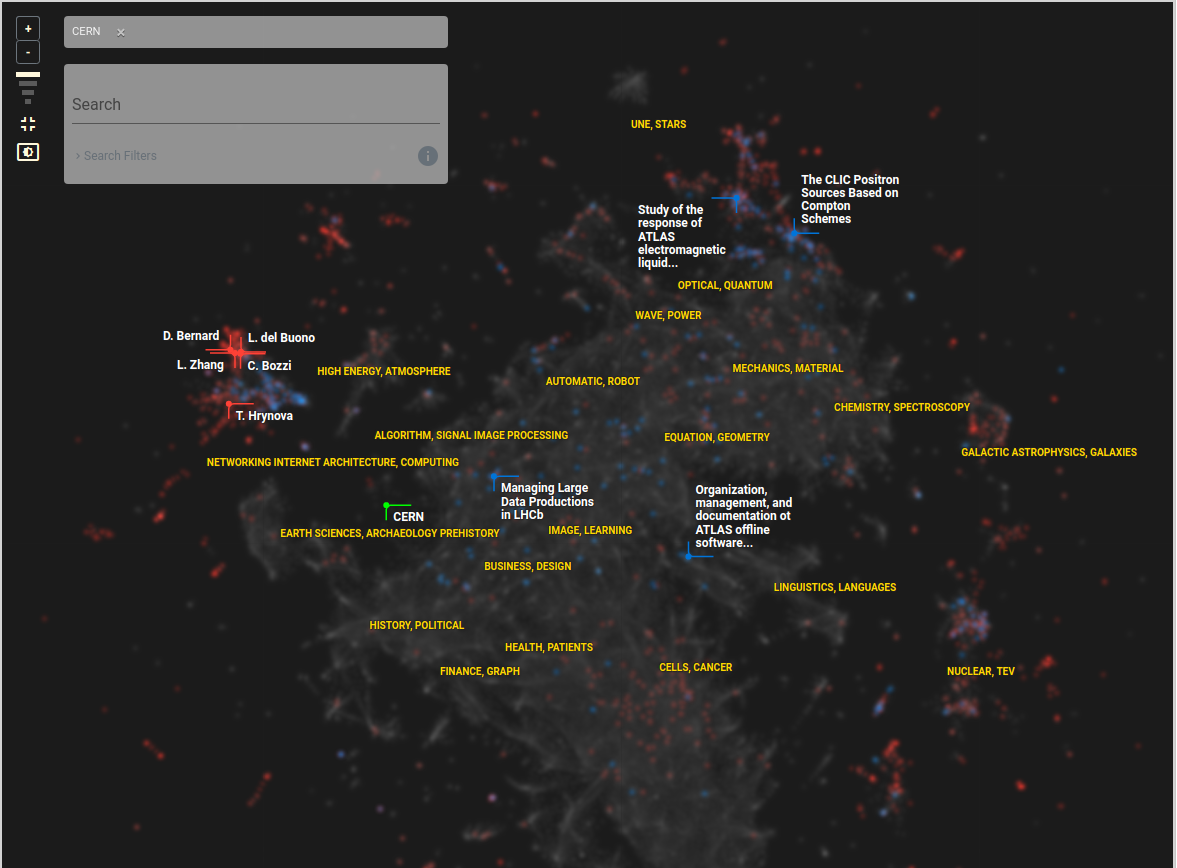}
  \caption{Dynamically filtering the density maps by content lets users explore specific contextual areas and visually check their scope. Here only articles and authors associated with the CERN are displayed while the rest of the map is grayed out.}
  \label{fig:cern}
\end{figure*}

\paragraph{Filters}
\label{subsec:filters}
While density maps efficiently aggregate high amounts of information, they make it hard to see details beyond the white and yellow landmarks.
To see the actual points in a region and get the details, users can specify a filter and only retain entities that match the filter (type of entity or property). For instance, using the filter ``CERN'', \autoref{fig:cern} shows only articles and authors linked to the CERN Physics Laboratory, greying-out other entities.
As intended, this filter makes it clear that the thematic areas most relevant to CERN are astrophysics and particle physics; other areas contain few, if any, publications or authors associated with CERN. 


The overall \Car\ pipeline is detailed in the next two sections, distinguishing \CarD\ and \CarV\ parts. 

%% file: 03-A-CartoData.tex
\begin{figure*}
  \centering
  \includegraphics[width=\linewidth]{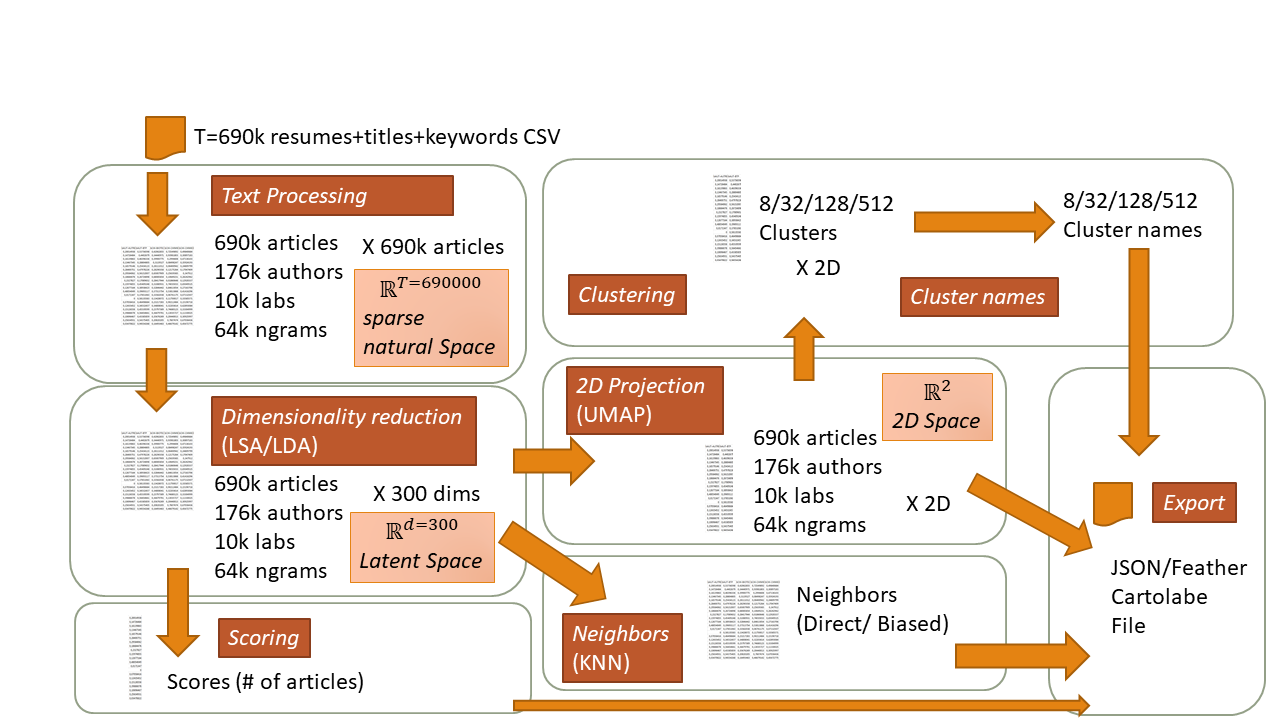}
  \caption{\CarD\ data processing pipeline illustrated on the HAL use case.}
  \label{fig:pipeline}
\end{figure*}

\section{\CarD\ Pipeline}
\label{subsec:pipeoverview}

The \CarD\ pipeline is illustrated on \autoref{fig:pipeline}. 
The first step is to collect the data, i.e., the document corpus. In the HAL application, the repository is queried through its open API, serving metadata on circa 800k scientific publications and  600k full texts in PDF format, including most (ideally all) articles funded by the French public research.

Note that the corpus involves different types of entities, e.g., articles, authors, laboratories, and institutes. The core of \CarD\ consists in homogeneously and efficiently transforming this textual corpus and all entities therein into a collection of 
points usable by the visualization engine \CarV. 

After pre-processing, 
the entities are mapped onto a high-dimensional space. 
Landscapes and clusters are formed and named 
to facilitate the further navigation of the map. \CarD\ eventually outputs a file storing the 2D coordinates of all the entities and their relationships that will be exploited by \CarV. 

\paragraph{Pre-processing}\label{subsec:preprocess}
Each article is described by its identifier, title, abstract, keywords, publication date, research domain, authors, and author institutions\rev{, each collected automatically from the HAL metadata}.

Entities include: words (more specifically \rev{\emph{n-grams},} terms made of sequences of at most $N=5$ words); articles (viewed as bags of words);  authors (viewed as bags of articles); and laboratories or institutes (viewed as bags of articles). Only the titles, abstracts and keywords of the articles are used at the moment\rev{; we can use the full text or any section that would be the most informative, but \revv{due to the lack of quality measures to determine which sections of the text are the most informative ones, we use the simplest parts:}\dell{lack of quality measure to assess which section of the text is the most informative, we use the easiest parts:} the title, the keywords and the abstract}. 
As of Dec. 2019, the HAL collection includes 800k articles, 200k authors, 10k laboratories, and 100k N-grams.

Articles undergo standard NLP pre-processing, including stop words removal 
and cleanup, only retaining terms with a minimum number $m=25$ of occurrences. This step also determines the size ($V = 64,000$) of the considered vocabulary.  Thereafter, a word matrix and an author matrix are computed.

The word matrix $M_{T,V}$ is a $T \times V$ matrix, with $T$ the number of articles and $V$ the vocabulary size. $M_{T,V}(i,j)$ is the term-frequency-inverse-document-frequency (tf-idf) of term $j$ in document $i$.
%
Likewise, the author (and laboratory) matrix $M_{T,L}$ is a $T \times L$ matrix, with $L$ the number of authors (resp. laboratories), where 
$M_{T,V}(i,k) = 1$ iff the $k$\ts{th} author (resp. laboratory) co-authored the $i$\ts{th} article. Only authors and laboratories with a minimum number of papers (set to 3 in the following) are considered.

\paragraph{Continuous Embedding}\label{sec:hd}
Besides their above primary representation, entities are attached a \emph{latent} representation:  a continuous embedding is used to map each entity onto a vector in $\RR^{d}$, with $d$ typically in the hundreds. 
This continuous embedding firstly reduces the dimensionality of the problem by two orders of magnitude, enhancing the computational tractability of our approach. Secondly and most importantly, it supports the definition of a sensible similarity among entities, avoiding the pathologies of Euclidean distance in high dimensional spaces, the so-called curse of dimensionality. 

Preliminary experiments were conducted considering Latent Semantic Analysis (LSA)~\cite{lsa} and Latent Dirichlet Allocation (LDA)~\cite{lda} with $d$ ranging from 80 to 500. In the following, we use LSA with $d=300$ \rev{that we found effective for HAL after several trials. Each article is therefore represented as a vector of 300 topics (the latent space) computed by LSA; authors are represented in the same space, as well as laboratories, teams, and institutions. }

\paragraph{Topology}
To each entity are associated its neighbors of each type, where the distance among two entities is their Euclidean distance in the latent space $\RR^d$. 
This topology supports a main \Car\ functionality, namely visual information retrieval. As explained, users might want to find articles \emph{similar} to an article, or relevant to a topic.
More generally, the topology of the latent space is good iff the user can make sense of the map, finding that similar items are  (most of the time) close to each other. 

For the sake of real-time querying, the $k$ nearest neighbors of each article are pre-computed ($k=10$); a fast approximate $k$-nn algorithm is used, with quasi-linear complexity in the number of articles. 

\paragraph{2D Projection}
\Car\ uses the \UMAP~\cite{2018arXivUMAP} multidimensional projection technique
for its good scalability and stability to project high-dimensional data in 2D, focusing on preserving close neighborhood relationships. 
Most interestingly, \UMAP\ enables to approximate the 2D projection in the case of very large datasets (Section ``Use Cases'').

\paragraph{Landmarks}
Navigating on the map is facilitated by defining and naming topics.
The construction of \del{intuitive}\rev{meaningful} cluster topics proceeds by using a \kmeans\ clustering on the articles and setting the distance of any two articles to their Euclidean distance in $\RR^2$ to avoid the issue of so-called \emph{stressed} entities. This issue reflects the fact that entities related to different themes (e.g., an author working on several topics) that are projected in different regions of the 2D map, will be located in between these regions, and thus far from each of them, hindering its interpretation. 

Topics are defined at 4 levels of granularity: 8 top-level, 24 medium-level clusters, 72 low-level clusters and 216 very low-level clusters, from the most general to the finest-grained topics. \rev{We selected these values experimentally, too-few would leave large areas unlabeled and too many would overcrowd the display.}
Each cluster is labeled after the most frequent word or term in the cluster (compared to the other clusters). If there exists no sufficiently frequent word in the cluster's articles, a second term (the most frequent one in the articles that do not contain the first term) is retained and both terms form the cluster name. A further requirement enforced is that two adjacent clusters should have different names. 

Cluster labels are located at the cluster center and displayed in yellow, \del{depending on}\rev{according to} the level of zoom of the viewport. 
Other landmarks, defined as the most \del{salient}\rev{important} entities (e.g., prolific authors and highly cited articles according to a score of interest, see below), are displayed in white next to their entity's position. They appear dynamically according to the viewport, the display density, and their relative score. 

\paragraph{Scoring}
\label{sec:scoring}
To each entity is attached a score, expectedly reflecting its interest for the user. The design of more sophisticated scores (e.g., the citation counts for articles) is typically 
domain-dependent. Note that the modular organization of our pipeline permits specific implementations of the scoring functionality depending on the available data and its semantics.
In the HAL application, the scores attached to authors and laboratories (respectively words) are their number of articles (resp. the number of articles they appear in). 
The score of 
\del{any article currently is set to 1, and will be} \rev{articles is set to the number of views}\del{or citation count in the next \Car's HAL version} \rev{in HAL}. 
Only the 10 entities with highest score appearing in the viewport are displayed.

\paragraph{Export}
\label{sec:export}
The output of \CarD\ is a \del{JSON}\rev{tabular} file that is used as input by the \CarV\ server. 
%
This file contains, for each entity: \emph{i}) its id, its label, its score, its 2D position, and its type (article, author, \ldots); \emph{ii}) its 10 nearest neighbors; \emph{iii}) the relevant metadata depending on the type of the entity (e.g., the HAL link for the articles).

%% file: 03-B-CartoVis.tex
\section{\CarV\ Pipeline}
\label{sec:visu}

As described above, \CarV\ is provided with the data structure generated by \CarD. Most generally, this data structure is made of a list of entities or points, where each point is attached to its main features and its 10 nearest neighbors in each category. 
The simplicity of this data structure is at the core of the \Car\ genericity, allowing it to handle diverse types of documents and data collections as long as they can be represented as a list of positioned entities. 

\paragraph{Visualization}
 \revv{The}\dell{A} main goal of \Car\ is to make sense of large scale data, up to tens of millions of entities. A first functionality thus is to provide a high-level visualization, displaying an overview of the collection. 
To do so, \CarV\ creates density map images, reflecting the density of the point distribution in the diverse regions of the 2D space.
This functionality is achieved using \rev{standard Python libraries: NumPy to compute 2D histograms, SciPy to apply a smoothing Gaussian blur, and Pillow to convert them into images.}\del{Datashader library (\url{http://datashader.org/}) that can easily aggregate millions of points into a single static image. }
Still, it is essential for users to interactively zoom, pan,
and filter the data according to their exploration. Thus, a multi-scale visualization is produced by \emph{i}) generating offline static images; \emph{ii}) tiling them and serving them interactively.
%
The tiled pyramid is created by building static images of each zoom level of the density maps \del{using the Datashader library} and splitting them into $256 \times 256$ pixel \rev{grey-scale} tiles. These tiles are sent by the web server application to the browser following the tiled image protocol of D3. \revv{We generate a different heatmap for each entity that we want to visualize (authors and articles for HAL). This allows the users to show either of them dynamically.}
\rev{\dell{We generate a different heatmap for each entity that we want to visualize, authors and articles for HAL, to allow users to show either of them dynamically.} The heatmaps, created as standard grey-scale PNG images, are combined by the browser using standard SVG filters to color and blend them. }

\paragraph{The Context}
\CarV\ enables users to look for a specific region or point of interest, through displaying contextual information and/or enabling direct search and filters (below).
The contextual information, characterizing the region currently displayed on the screen, consists of 
labels superimposed on  top of the tiled density maps. For the sake of readability, only the top-scored entities and most general clusters located on the viewport are made visible (see ``Scoring'' Section).
\rev{We chose a yellow color for the cluster labels to limit their saliency since they remain displayed all the time and are only useful as area landmarks, we do not want them to attract the reader's attention as much as other landmarks. \revv{To be fully readable, the popular entities'  labels appear in white with a black outline. This increases their contrast and they stand out independently of their background.}\dell{The popular entities'  labels appear in white because they should be fully readable; to increase their contrast, they have a black outline so they stand out independently of their background.}}
\rev{Instead of outlining the labels with a colored border according to their type, we only draw a colored ``wedge'' in a corner of the label bounding box, pointing to the exact position of the label (see \autoref{fig:hal}). This encoding, similar to a glyph, saves the visualization from becoming overcrowded and is easily understood by users.}

%
\CarV\ relies on a standard database
to efficiently select the most \rev{important}\del{salient} entities based on their score and position to be \del{displayed}\rev{labeled} on the current viewport. 

\paragraph{Contextual Filters}
Besides panning and zooming, users can tailor the map to their current goals by using filters, e.g.,
showing only articles and authors associated with a targeted laboratory (the CERN in \autoref{fig:cern})
and greying out the others. 
In the HAL application, filters can specify the desired laboratory names (HAL references 10,000 of them) and the publication period (from 2008 to now). Filters can be combined, e.g., enabling to visualize the entities associated with the CERN or INRIA laboratories and published in 2018.

As pre-computing density maps for all possible filter masks is not feasible, 
contextual filtering proceeds by generating density maps upon users' query in a progressive way~\cite{Progressive}. Specifically, \CarV\ uses the Roaring Bitmaps library~\cite{Lemire2018} to create and store compressed indices of the entity identifiers associated with each elementary filter value.
%
\CarV\ thus is provided with a bitmap pre-computed for each map tile, storing all entities located in the tile region, and a bitmap for each filter value (lab and year), storing all ids of all articles matching this filter value.   
All bitmaps are combined interactively  in a few milliseconds 
to retrieve the identifiers of all articles matching the contextual filter, e.g., the CERN or INRIA articles published in 2018 and falling in the displayed tiles.  

Eventually, the selected entities are superimposed on the current density map image. \del{As the Datashader}\rev{Our} library is quite efficient at generating density images, this last operation can be done in reasonable time ($<$ 500ms), creating a tiled portion of the density map matching the users' criteria that is displayed on top of the shaded overall map. This process is done progressively, using a thin red progress bar on top of the screen as visual feedback indicating that the operation is ongoing during the 500ms to 2s of its duration, taking into account the time to send images through the network. 

Progressive computation and rendering is repeated each time the user zooms and pans the map, with controlled caching to improve performance.  This progressive generation of a multi-scale filtered density maps is a unique feature of \CarV. It opens new perspectives for implementing scalable web-based visualizations with dynamic queries.

%% file: 06-use-cases.tex
\section{Use Cases}\label{sec:usecases}

\begin{table*}[htbp]
\centering
\begin{tabular}{|l|l|r@{ }l|r@{ }r|} 
 \hline
 \multicolumn{1}{|c|}{\bf Dataset} & \multicolumn{1}{c|}{\bf Documents} & \multicolumn{2}{c|}{\bf Size} & \multicolumn{2}{c|}{\bf Computing times}\\
 \hline
 \multirow{5}{8em}{HAL} & \multirow{5}{14em}{Scientific publications} & & & Term extraction:& 360 s \\ 
 && 690,307 & Articles & LSA (300 components):& 2,489 s \\
 && 176,296 & Authors & UMAP:& 5,359 s \\
 && 65,053 & Terms & Clustering:& 338 s \\
 && 10,699 & Laboratories & Nearest neighbors:& 25,994 s \\
 \hline
 \multirow{5}{8em}{Wikipedia} & \multirow{5}{14em}{Encyclopedia entries}  &&& Term extraction:& 26,152 s \\
 && 4,631,475 & Articles & LSA (300 components):& 13,716 s\\
 && 200,000 & Terms & UMAP:& 13,869 s\\
 &&&& Clustering:& 13,898 s \\
 &&&& Nearest neighbors:& 140,334 s \\
 \hline
 \multirow{5}{8em}{Le Grand Débat} & \multirow{5}{14em}{Citizen propositions}  &&& Term extraction:& 6,073 s  \\
 && 4.300,000 & Propositions & LSA (300 components):& 14,338 s \\
 && 600,000 & Terms & UMAP:&  3,701 s\\
 &&&& Clustering:&  19,240 s\\
 &&&& Nearest neighbors:&  11,903 s\\
 \hline
\end{tabular}
\caption{Three \Car\ Use Cases: Datasets and Computational Time to process them.}
\label{table:usecases}
\end{table*}

Several use cases of \Car\ are described to illustrate its flexibility (\autoref{table:usecases}): basically all key components presented in the previous sections can be replaced with other Python libraries to accommodate other data structures or needs. Interestingly, new algorithms (achieving e.g., word embedding, 2D projection or clustering) can also be integrated into the \Car\ framework for comparative assessment with the state of the art\rev{: \Car\ is designed to be agnostic about the methods used to generate its data and flexible enough to accommodate most of them. \CarD\ comes with multiple pipelines, including the use cases described below. It also includes examples using different topic modeling methods such as LSA and LDA (used for all the production pipelines). We also provide example pipelines using word embedding methods such as doc2vec
and even an example using the MNIST hand-recognition dataset made of images instead of texts to showcase the flexibility of our framework.}

\paragraph{Wikipedia}
We used the Wikipedia dataset as a scalability study, confirming that \Car\ can manage millions of documents with no performance degradation. The main difference compared to the HAL application regards the \UMAP\ projection: when building a map of all articles in the English Wikipedia, 
an approximated projection was trained from a uniform sample of 20\% of the articles and used for all articles, to preserve the projection performance.
Another difference is that a single density map, reflecting the articles and based on the top 200k words, was built instead of two for HAL (one for the articles and one for the authors). 
Lastly, the scoring functionality was modified to account for the article popularity in Wikipedia, first displaying the most popular articles in the viewport.

\paragraph{Le Grand Débat}
\label{sec:granddebat}
The {\em Grand Débat} dataset includes the citizens' full-text contributions in response to the French government questions about  societal and political issues. This {\em Grand Débat} initiative took place in France from January to March 2019 (\http{granddebat.fr}), gathering circa 4M full-text contributions. \Car\ was successfully used to explore this large and heterogeneous corpus upon an official request from a French regulation organization, to check if the eventual analysis and summary of these contributions were fair and unbiased in terms both of topic coverage and summary contents. 

Like for Wikipedia, the real-time performance is enforced by 
approximating the \UMAP\ projection, using a projection model learned from a subset of the data. The filtering options include each one of the 600k extracted terms, supporting the full-text querying of the dataset.

%% file: 07-discussion.tex
\begin{revs}
\section{Evaluation}

We conducted a \dell{IRB approved} usability study \revv{approved by our Institutional Review Board} with 7 users from our laboratory (face to face before the COVID-19 outbreak) on the HAL dataset. \revv{This preliminary study was found insightful, and it will be continued to improve}\dell{The study was preliminary but insightful and will be continued to improve} \Car. We had three goals:\revv{\ }1) Verifying the usability of \CarV\ for our target users; 2) Verifying the quality of \CarD\ regarding the distance measure and its projection; and 3) Collecting feedback on possible improvements regarding both \CarD\ and \CarV.

\noindent\textbf{Demographic} The 7 users (2 female, 5 male, age range 30--60) were researchers (2 full professors, 4 assistant professors, 1 researcher). They were recruited by sending an email to the laboratory mailing list. None of them were familiar with visualization or projection methods, and they did not receive any compensation.

\noindent\textbf{Setup} The study lasted about one hour. Each user started by reading a consent form explaining the goal of the study, with the ability to ask questions to the interviewer, and then signing the form. Each session was split in four parts: 1) Training; 2) Usability testing; 3) Assessing the metrics and its visualization; and 4) Collecting feedback. \revv{Participants filled a questionnaire during and after the study to keep track of their answers and comments.}

\noindent\textbf{Training} We introduced the main features of \CarV\ and its UI, starting with the visible features without interaction, then the possible interactions on each of the visualized items and UI areas, explaining the function of each button, text area, and icons. Then\revv{,} the search capability was explained with several examples. 

\noindent\textbf{Usability Testing} We asked each user to perform a standardized set of tasks to make sure they understood the training well enough and were able to use \Car\ by themselves.

\noindent\textbf{Assessing the metrics and its visualization} We designed a five steps protocol to assess the quality of the distance metrics and its visualization. 1) We asked users to list the names of french authors well-known in their area of competence, not using any tool. 2) We then asked them to find the authors using \Car\ and to comment on the cluster names while navigating around; 3) We asked if, in light of the results, \Car\ appeared trustworthy. Then, we asked to repeat the same task (gather names and check with the tool) for other domains or fields they were familiar with, and report on their findings. 4) We asked to assess the proximity of articles and authors, starting from a recent article well known by the user, select a list of authors related to it and check if they were close in the visualization. 5) We asked to select three PhD students and to assess the accuracy of their clusters and their neighborhoods.

\noindent\textbf{Collecting feedback} Users were asked if they wished to continue \revv{exploring}\dell{explore} the map, and then to report at a higher level on what they liked, what they found useful and less useful, difficult, and about the utility of the tool.

\subsubsection{Results}

We gathered a total of 180 comments, 2/3 were about the UI and visualization\revv{,} and 1/3 about the metrics. 

\noindent\textbf{Usability Testing} All the users were able to complete all the tasks we asked them to do after the training, validating the usability of \CarV\ for the selected tasks.

\noindent\textbf{Assessing the metrics and its visualization} Due to limited space, we only report on the results on steps 1--3. Users have listed a total of 28 names at step 1, representative of their domain. 25 have been found on the map at step 2 (90\%), 3 were not found, and 2 were found at unexpected positions (missing neighbors artifact from the \UMAP\ projection). During step 3, our users have collected 53 names around their areas of interest on the map (including their ground truth), none of which were considered as obvious mistakes (no false neighbors). Therefore, they all considered \Car\ trustworthy.

All the users who looked at the 10 closest n-grams to an article or an author were satisfied by the results, meaning that the topic modeling results were good for our small sample.

All the users have been able to perform the tasks on metrics and visualization as expected by our protocol and to answer all our questions without major problem, witnessing that \Car\ is already usable in its current form for neighborhood exploration tasks, and that the NLP and projections were fair to the data. 

\noindent\textbf{Collecting feedback} 
The main UI issues raised were: 1) improve the icons and their affordance when they can be used to toggle some features, such as hiding/showing layers under the visualization, 2) add bookmarks to keep track of and retrieve interesting views, 3) improve the highlighting of selected features \revv{(selected labels and points not salient enough)}, and 4) improve the topic clusters labels (shown in yellow) and their stability during the navigation.
We are working on addressing these usability issues but were pleased by these initial usability results. We even received an unexpected use case from a PhD student, who realized that the list of well-known authors close to his articles gave him hints on where to apply for a position.

\medskip

Overall, this preliminary study has validated our design goals. We still need to continue our evaluation with more users, but our protocol seems valid and we will continue to use it to further improve \Car.
\end{revs}

\section{Scalability}

\Car\ is scalable to circa ten million documents: it was specifically designed to handle large datasets. While the text processing pipeline might take time to complete, this is an offline process that does not hinder \Car\ interactive system performance.  The use cases described in the previous section demonstrate that \Car\ can display millions of points in a browser without any visible slowdown. The use of static images as density maps is not affected by the size of the data. Though \Car\ relies on an external database to index the points and query them to display labels in the viewport, the scalability of existing databases is well documented and can sustain the browsers' demand.  
The use of the Roaring Bitmap library to build indices for the filter function also offers high scalability and performs well even with millions of points and filter values.

\rev{As for the human side, we believe \Car's visualization can scale to tens of millions of documents for overview tasks using its heatmap. The labeling and selection mechanisms rely on a good scoring for items that is sometimes difficult to obtain for large datasets. We have been able to get it for HAL with difficulties, Wikipedia provides its access statistics, but arXiv does not.  Without these scores, there is no simple way to label the most important items in the visualization and the labels shown are random samples instead of real landmarks. 
When selecting an author, an article, or a word, we only allow displaying its ten closest neighbors. Some users complained about it and for larger datasets, we will have to increase this number, but this is currently a parameter set in \CarD\ and in \CarV\ and it can already be changed.  More user studies will inform us of possible issues for larger \revv{datasets}\dell{dataset}.}

%% file: 08-conclusion.tex
\section{Conclusion and Perspectives}

The \Car\ framework is a web-based system aimed at visualizing and exploring large document collections for 
users interested in these documents but not familiar with visualization and machine learning. This framework, initially designed to explore a scientific publication corpus, has evolved towards a generic system accommodating different document collections. Written in Python, it is freely available online.

The \Car\ design relies on a multi-scale 2D representation (map) that allow to make sense of large-scale data through an overview, and panning and zooming navigation. In addition, multiple interactive features allow to search, filter, and explore the neighborhoods of selected items or regions.

Our next step will be to launch a call to gather new corpora to further test the genericity of \Car, to perform more exhaustive evaluations of the multiple stages of the system, and to apply it to new interesting corpora needing exploration by topics such as libraries, archives, and software repositories.


%% file: cartolabe.bbl
\begin{thebibliography}{10}
\providecommand{\url}[1]{#1}
\csname url@samestyle\endcsname
\providecommand{\newblock}{\relax}
\providecommand{\bibinfo}[2]{#2}
\providecommand{\BIBentrySTDinterwordspacing}{\spaceskip=0pt\relax}
\providecommand{\BIBentryALTinterwordstretchfactor}{4}
\providecommand{\BIBentryALTinterwordspacing}{\spaceskip=\fontdimen2\font plus
\BIBentryALTinterwordstretchfactor\fontdimen3\font minus
  \fontdimen4\font\relax}
\providecommand{\BIBforeignlanguage}[2]{{%
\expandafter\ifx\csname l@#1\endcsname\relax
\typeout{** WARNING: IEEEtran.bst: No hyphenation pattern has been}%
\typeout{** loaded for the language `#1'. Using the pattern for}%
\typeout{** the default language instead.}%
\else
\language=\csname l@#1\endcsname
\fi
#2}}
\providecommand{\BIBdecl}{\relax}
\BIBdecl

\bibitem{lsa}
S.~Deerwester, S.~T. Dumais, G.~W. Furnas, T.~K. Landauer, and R.~Harshman,
  ``{Indexing by latent semantic analysis},'' \emph{Journal of the American
  Society for Information Science}, vol.~41, no.~6, pp. 391--407, 1990.

\bibitem{lda}
D.~M. Blei, A.~Y. Ng, and M.~I. Jordan, ``{Latent Dirichlet Allocation},''
  \emph{Journal of machine Learning research}, vol.~3, no. Jan, pp. 993--1022,
  2003.

\bibitem{Nonato2018}
L.~G. Nonato and M.~Aupetit, ``Multidimensional projection for visual
  analytics: Linking techniques with distortions, tasks, and layout
  enrichment,'' \emph{{IEEE} Transactions on Visualization and Computer
  Graphics}, pp. 1--1, 2018.

\bibitem{Progressive}
J.-D. Fekete, D.~Fisher, A.~Nandi, and M.~Sedlmair, ``{Progressive Data
  Analysis and Visualization (Dagstuhl Seminar 18411)},'' \emph{Dagstuhl
  Reports}, vol.~8, no.~10, pp. 1--40, 2019.

\bibitem{ThemeScapes}
J.~A. Wise, J.~J. Thomas, K.~Pennock, D.~Lantrip, M.~Pottier, A.~Schur, and
  V.~Crow, ``Visualizing the non-visual: Spatial analysis and interaction with
  information for text documents,'' in \emph{Readings in Information
  Visualization}, S.~K. Card, J.~D. Mackinlay, and B.~Shneiderman, Eds.\hskip
  1em plus 0.5em minus 0.4em\relax San Francisco, CA, USA: Morgan Kaufmann
  Publishers Inc., 1999, pp. 442--450.

\bibitem{Kucher2015}
\BIBentryALTinterwordspacing
K.~Kucher and A.~Kerren, ``{Text visualization techniques: Taxonomy, visual
  survey, and community insights},'' in \emph{2015 {IEEE} Pacific Visualization
  Symposium ({PacificVis})}.\hskip 1em plus 0.5em minus 0.4em\relax IEEE, apr
  2015. [Online]. Available:
  \url{https://doi.org/10.1109/pacificvis.2015.7156366}
\BIBentrySTDinterwordspacing

\bibitem{DBLP:conf/apvis/FriedK14}
\BIBentryALTinterwordspacing
D.~Fried and S.~G. Kobourov, ``Maps of computer science,'' in \emph{{IEEE}
  Pacific Visualization Symposium, PacificVis 2014, Yokohama, Japan, March 4-7,
  2014}, 2014, pp. 113--120. [Online]. Available:
  \url{https://doi.org/10.1109/PacificVis.2014.47}
\BIBentrySTDinterwordspacing

\bibitem{vosviewer}
N.~van Eck and L.~Waltman, ``Software survey: {VOSviewer}, a computer program
  for bibliometric mapping,'' \emph{Scientometrics}, vol.~84, pp. 523--538,
  2010.

\bibitem{Espadoto2019}
\BIBentryALTinterwordspacing
M.~Espadoto, R.~M. Martins, A.~Kerren, N.~S.~T. Hirata, and A.~C. Telea,
  ``{Towards a Quantitative Survey of Dimension Reduction Techniques},''
  \emph{{IEEE} Transactions on Visualization and Computer Graphics}, 2019.
  [Online]. Available: \url{https://doi.org/10.1109/tvcg.2019.2944182}
\BIBentrySTDinterwordspacing

\bibitem{2018arXivUMAP}
L.~{McInnes}, J.~{Healy}, and J.~{Melville}, ``{UMAP: Uniform Manifold
  Approximation and Projection for Dimension Reduction},'' \emph{ArXiv
  e-prints}, Feb. 2018.

\bibitem{UTOPIAN}
J.~Choo, C.~Lee, C.~K. Reddy, and H.~Park, ``{UTOPIAN: User-Driven Topic
  Modeling Based on Interactive Nonnegative Matrix Factorization},''
  \emph{{IEEE} Transactions on Visualization and Computer Graphics}, vol.~19,
  no.~12, pp. 1992--2001, Dec 2013.

\bibitem{BioVisExplorer}
\BIBentryALTinterwordspacing
A.~Kerren, K.~Kucher, Y.-F. Li, and F.~Schreiber, ``{BioVis Explorer: A visual
  guide for biological data visualization techniques},'' \emph{{PLOS} {ONE}},
  vol.~12, no.~11, p. e0187341, Nov. 2017. [Online]. Available:
  \url{https://doi.org/10.1371/journal.pone.0187341}
\BIBentrySTDinterwordspacing

\bibitem{nanocubes}
L.~Lins, J.~T. Klosowski, and C.~Scheidegger, ``{Nanocubes for Real-Time
  Exploration of Spatiotemporal Datasets},'' \emph{{IEEE} Transactions on
  Visualization and Computer Graphics}, vol.~19, no.~12, pp. 2456--2465, 2013.

\bibitem{Lemire2018}
D.~Lemire, O.~Kaser, N.~Kurz, L.~Deri, C.~O'Hara, F.~Saint-Jacques, and
  G.~Ssi-Yan-Kai, ``Roaring bitmaps: Implementation of an optimized software
  library,'' \emph{Software: Practice and Experience}, vol.~48, no.~4, pp.
  867--895, jan 2018.

\end{thebibliography}
